\documentclass[aps,prd,reprint,superscriptaddress,nofootinbib,nobibnotes,notitlepage]{revtex4-2}

\pdfoutput=1

\usepackage[T1]{fontenc}
\usepackage{aas_macros}   
\usepackage{graphicx}
\usepackage{amsmath}
\usepackage{amssymb}
\usepackage{bm}
\usepackage{xcolor}
\usepackage[colorlinks=true,linkcolor=blue!60!black,citecolor=blue!60!black,urlcolor=blue!60!black]{hyperref}
\usepackage[capitalise,nameinlink]{cleveref}
\Crefname{figure}{Fig.}{Figs.}
\Crefname{equation}{Eq.}{Eqs.}
\Crefname{table}{Table}{Tables}
\Crefname{appendix}{Appendix}{Appendices}
\crefname{appendix}{Appendix}{Appendices}

\newcommand{\bk}{\bm{k}}

\newcommand{\fnl}{f_{\rm NL}}
\newcommand{\fnll}{f_{\mathrm{NL}}^{\mathrm{loc}}}
\newcommand{\fnle}{f_{\mathrm{NL}}^{\mathrm{equil}}}
\newcommand{\fnlo}{f_{\mathrm{NL}}^{\mathrm{orth}}}
\newcommand{\Mpc}{\ensuremath{h\,\mathrm{Mpc}^{-1}}}
\newcommand{\kmax}{k_{\rm max}}
\newcommand{\kmin}{k_{\rm min}}
\newcommand{\sigv}{\sigma_v}
\newcommand{\kf}{k_{\rm f}}
\newcommand{\Mh}{M_{\rm h}}

\graphicspath{{./plots/}}

\begin{document}

\title{Constraining Inflation with Little Red Dots}

\author{Dionysios Karagiannis}
\email{dakaragian@gmail.com}
\affiliation{Van Swinderen Institute for Particle Physics and Gravity, University of Groningen, 9747 AG Groningen, The Netherlands}
\affiliation{Department of Physics \& Astronomy, University of the Western Cape, Cape Town 7535, South Africa}

\author{P.~Daniel~Meerburg}
\affiliation{Van Swinderen Institute for Particle Physics and Gravity, University of Groningen, 9747 AG Groningen, The Netherlands}

\begin{abstract}
Primordial non-Gaussianity is a direct probe of the physics of inflation.
The tightest bounds still come from the cosmic microwave background, but
large-scale structure is rapidly catching up: its intrinsically
three-dimensional maps contain far more modes, provided the tracer combines
large volume, high redshift, strong bias and sufficient number density. The
Little Red Dots (LRDs) discovered by JWST plausibly offer all four. Inspired by a recent study on BAO measurements using LRDs, we forecast
the joint galaxy power spectrum and bispectrum of a $14{,}000\,{\rm deg}^2$ LRD
spectroscopic survey over $4<z<9$, marginalising over $32$ nuisance parameters
and including the non-Gaussian bispectrum covariance. We find
$\sigma(\fnll)=0.32$, $\sigma(\fnle)=32$ and $\sigma(\fnlo)=10$, improving on
{\em Planck} by factors of $16$, $1.5$ and $2.3$. The local bound lies a factor of
three below the $\sigma(\fnll)\simeq1$ threshold that separates broad classes of
single- and multi-field inflation. LRD clustering therefore has the potential to
become an important probe of inflation, and makes a strong case for future
wide-field near- to mid-infrared spectroscopy.
\end{abstract}

\maketitle

\section{Introduction}

Inflation is the leading scenario for the early Universe, and one of its great
successes is that it provided an explanation for the origin of cosmic structure. The physics, mainly the fields and
interactions that drove it, however, remain unknown. Primordial non-Gaussianity
(PNG) provides a direct observational route to this physics: its amplitude and
momentum dependence distinguish additional fields, non-trivial interactions and
departures from the standard
vacuum~\cite{Maldacena2003,Bartolo2004,Komatsu2009,Achucarro_2022}, and may test
the quantum origin of the primordial fluctuations~\cite{Green_2020}. For weak
non-Gaussianity the leading observable is the primordial bispectrum,
conventionally described by local, equilateral and orthogonal templates.
Reaching $\sigma(\fnll)<1$ would cross a decisive threshold between single- and
multi-field models of
inflation~\cite{dePutter:2016trg,Meerburg:2019qqi,Achucarro_2022}, shedding light on these early stages of the Universe.

The cosmic microwave background (CMB) has provided the tightest constraints,
with {\em Planck} finding $\sigma(\fnll)=5.1$, $\sigma(\fnle)=47$ and
$\sigma(\fnlo)=24$~\cite{Planck:2019kim} and future constraints are projected to improve the constraints by a factor of a few~\cite{SimonsObservatory:2018koc,LiteBIRD:2022cnt}. On the other hand, large-scale-structure (LSS) surveys
are now rapidly catching up\footnote{CMB constraints are expected to remain leading for at least a while for non-local shapes.}. By mapping structure in three dimensions they
access many more primordial modes than the projected CMB sky, and local PNG
additionally imprints a distinctive scale-dependent tracer
bias~\cite{Dalal2008,Slosar2008,Matarrese2008,Desjacques2010,Desjacques2016},
which gives the power spectrum direct sensitivity to $\fnll$, complementary to that of the bispectrum. Present LSS constraints nonetheless remain weaker than
{\em Planck}~\cite{Castorina:2019wmr,Mueller:2021jbt,Cabass:2022wjy,Cabass:2022ymb,
DAmico:2022gki,Cagliari_2023,Chaussidon_2024}, because many of the additional
modes lie in the nonlinear regime, where the primordial signal must be separated
from the much larger bispectrum component generated by gravitational
evolution.\footnote{Since nonlinear evolution acts locally, it cannot generate
the long-range correlations that carry the primordial signal, which can be
exploited to break this degeneracy~\cite{Baumann:2021ykm}.}

The way forward is therefore determined by the potential of the observable tracer field. Exploratory forecasts across the
observational spectrum, from optical and near-infrared galaxy
surveys~\cite{Spherex_2015,Karagiannis2018,Sailer:2021yzm,Heinrich:2023qaa,Euclid:2025hlc} to radio 21\,cm
intensity mapping~\cite{Fonseca:2015laa,Camera:2014bwa,Karagiannis:2019jjx,Karagiannis:2020dpq,Karagiannis2026bimodal}, consistently
favour the same combination: a sample spanning a large volume at high redshift,
where more modes remain perturbative, with a large bias and sufficient number
density to overcome the suppressed matter clustering and shot
noise~\cite{Meerburg:2019qqi,MoradinezhadDizgah2020}. A large volume is especially
valuable for local PNG, giving access to the most squeezed triangles, where the
signal is largest. The third dimension also matters: projecting the primordial
field onto a two-dimensional surface, as in the CMB, blurs short-distance modes
and degrades the signal-to-noise scaling of equilateral-like shapes much more
than that of squeezed ones~\cite{Kalaja:2020mkq}, a loss that three-dimensional
LSS maps avoid.

The Little Red Dots (LRDs) discovered by JWST~\cite{Matthee2024} appear to
combine these properties. These compact, red, broad-line sources are most
numerous at $4\lesssim z\lesssim9$~\cite{Kokorev24,kocevski2025rise,Labbe2025,
Tanaka25,Park26,Ma26,Rinaldi26}, with number densities
$\bar n\sim10^{-4}\,h^3{\rm Mpc}^{-3}$ and a linear bias rising from
$b_1\simeq3.4$ to $\simeq8$ across this
interval~\cite{sun2026little,lin2026large,Inayoshi2025}. Their physical nature
remains debated~\cite{Naidu25,degraaff25pop,Inayoshi25rev,Madau26}, but their
value as an LSS tracer depends only on measurable properties: abundance, bias
and a distinctive spectrophotometric signature that enables efficient redshift
measurements~\cite{Hviding25,Setton25,Matthee26}. This combination has already
made LRDs a promising matter-era baryon acoustic oscillation
probe~\cite{Zebrowski2026}; here we show that it also makes them a powerful
probe of inflation.

In this paper we forecast the joint galaxy power spectrum and bispectrum of a
$14{,}000\,{\rm deg}^2$ LRD spectroscopic survey. We generate the complete bias
hierarchy from a single halo occupation model calibrated to the measured LRD
stellar mass, and include the non-Gaussian bispectrum covariance. Our main
results suggest that a targeted LRD survey would improve on {\em Planck} for all three templates, and would cross
the $\sigma(\fnll)=1$ threshold by a factor of three. Crucially, the non-Gaussian
covariance weakens the local constraint by a factor of four to six relative to
the Gaussian diagonal approximation, but in doing so renders it nearly
insensitive to the small-scale cut.

\section{The LRD sample}
We adopt the survey definition of Ref.~\cite{Zebrowski2026}: a DESI-like
$14{,}000\,{\rm deg}^2$ footprint and four wide redshift bins covering $4<z<9$,
with comoving number densities
$\bar n=(0.6$--$1.6)\times10^{-4}\,h^3{\rm Mpc}^{-3}$ taken from the abundance
model of Ref.~\cite{Inayoshi2025}, which is calibrated on the LRD samples of
Refs.~\cite{kocevski2023hidden,kocevski2025rise}, and a total comoving volume of
$190\,(\mathrm{Gpc}/h)^3$. We assume spectroscopic redshifts with
$\sigma_z/(1+z)=10^{-3}$ and fix the background cosmology to {\em Planck}
2018~\cite{Planck2018}. The per-bin properties are collected in
\Cref{tab:sample}.

The full bias hierarchy is generated
from a halo occupation distribution (HOD), so that $b_1$, $b_2$, $b_{s^2}$ and
the PNG responses all follow from a single occupation function. LRDs are taken to be central objects with an occupation
that is lognormal in host mass, of width $\sigma_{\rm SHMR}=0.30$\,dex and
centred on the host halo mass obtained by holding the LRD stellar mass at the
measured $\log_{10}(M_\star/M_\odot)=8.3$~\cite{sun2026little} and inverting
the UniverseMachine stellar-to-halo mass relation
(SHMR)~\cite{behroozi2019universemachine}; the host mass decreases from
$\log_{10}(\Mh/M_\odot)=11.03$ at $z=4$ to $10.72$ at $z=9$, consistent with the
direct clustering estimates of Ref.~\cite{lin2026large}. The SHMR sets the
distribution of host masses, and hence of biases, while the abundance only fixes the
fraction of occupied haloes (the duty cycle).
Convolving the occupation with the Tinker mass function and halo
bias~\cite{Tinker2008,tinker2010large} and using the standard co-evolution and
peak-background-split
relations~\cite{Baldauf2012,Chan2012,Desjacques2016} gives a
linear bias rising from $b_1=3.4$ to $8.0$ across the four bins, and a local-PNG
response $b_\phi$ rising from $8.1$ to $23.7$; the remaining biases are listed in
\Cref{tab:sample} and the occupation model and bias expressions are given in
\cref{app:bias}.

The predicted linear bias reproduces the results of Ref.~\cite{Zebrowski2026} to
better than $1\%$, indicating that the occupation model and the tabulated
abundance are mutually consistent; tuning the occupation to reproduce the
tabulated $b_1$ of Ref.~\cite{Zebrowski2026} directly, rather than predicting
it, changes $\sigma(\fnl)$ by less than $2\%$ for every shape and scale cut. The PNG responses assume a universal halo
mass function; since this need not hold for a selected tracer, we test this
assumption below.

\section{Forecast setup}

We quantify the constraining power of the LRD sample with the information
matrix formalism, applied to the redshift-space galaxy power spectrum
$P(\bk,z)$ and bispectrum $B(\bk_1,\bk_2,\bk_3,z)$. In a
given redshift bin the power spectrum information matrix is
\begin{equation}
F^{P}_{\alpha\beta}(z)=\int_{-1}^{1}\!\frac{{\rm d}\mu}{2}
\sum_{k=\kmin}^{\kmax}
\frac{1}{{\sf C}^{P}(k,\mu)}\,
\frac{\partial P(\bk)}{\partial\theta_\alpha}\,
\frac{\partial P(\bk)}{\partial\theta_\beta}\,,
\label{eq:fisherP}
\end{equation}
where $\mu=\hat\bk\cdot\hat{\bm z}$ and the Gaussian variance of a $k$-shell of
width $\Delta k$ is
${\sf C}^{P}=2(2\pi)^3 P^2(\bk)/(V_{\rm s}V_k)$, with $V_k=4\pi k^2\Delta k$ and
$V_{\rm s}$ the volume of the bin. For the bispectrum,
\begin{equation}
F^{B}_{\alpha\beta}(z)=\int\!{\rm d}^3\bk\,
\frac{\partial B(\bk_i)}{\partial\theta_\alpha}\,
\big[{\sf C}^{-1}\big]_{ij}\,
\frac{\partial B(\bk_j)}{\partial\theta_\beta}\,,
\label{eq:fisherB}
\end{equation}
where $\bk_i$ abbreviates the three sides of the $i$th triangle,
${\sf C}_{ij}$ is the triangle covariance discussed below, and
$\int{\rm d}^3\bk\equiv(1/4\pi)\int_{-1}^{1}{\rm d}\mu_1\int_0^{2\pi}{\rm d}\phi
\sum_T$, with $\mu_1$ and $\phi$ the polar and azimuthal orientation of the
triangle relative to the line of sight. The sum $\sum_T$ runs over all closed
triangles with $\kmin\le k_3\le k_2\le k_1\le\kmax$, binned with the survey
fundamental mode $\kf$ to resolve the squeezed configurations (see \cref{app:binning}). The
total information is
$\tilde F_{\alpha\beta}=\sum_{z_i}[F^{P}_{\alpha\beta}(z_i)+F^{B}_{\alpha\beta}(z_i)]$,
treating the four redshift bins as independent, and the marginalised error is
$\sigma(\theta_\alpha)=(\tilde F^{-1}_{\alpha\alpha})^{1/2}$. We neglect the
power spectrum--bispectrum cross-covariance, whose effect on the joint PNG
constraints is minor over the scales considered
here~\cite{Barreira:2019icq,Biagetti:2021tua}.

The model entering \Cref{eq:fisherP,eq:fisherB} is the tree-level redshift-space
power spectrum and bispectrum of a biased tracer, including the local-PNG
contributions to the first- and second-order kernels, infrared
resummation~\cite{Assassi2015,Ivanov:2021kcd}, Fingers-of-God and
redshift-error damping in both statistics, and the Alcock--Paczy\'nski
distortion~\cite{Alcock1979,Seo2003}; the expressions are collected in \cref{app:model}.
Each redshift bin carries eleven free parameters -- $H(z)$, $D_A(z)$, $f(z)$,
$b_1$, $b_2$, $b_{s^2}$, the two velocity-dispersion scales $\sigma_P$ and
$\sigma_B$, and three stochastic amplitudes $P_\varepsilon$,
$P_{\varepsilon\varepsilon_\delta}$ and $B_\varepsilon$ -- of which the
geometric and growth quantities are projected onto the cosmological basis
$\{\omega_b,\omega_c,h,n_s,\ln A_s\}$, leaving $32$ marginalised nuisance
parameters in total. All errors on $\fnl$ quoted below are marginalised over the
five cosmological parameters and the $32$ nuisance parameters, and $\fnl$ is
given in the CMB convention \cite{Camera:2014bwa}.

To assess how the forecast depends on the scales included, we consider two
small-scale cuts tied to the linear velocity dispersion $\sigv(z)$,
$\kmax(z)=\alpha/\sigv(z)$: a conservative cut, $\alpha=0.50$, which we adopt
for our headline results, and an optimistic cut, $\alpha=0.75$. At these
redshifts they correspond to $\kmax=0.37$--$0.60$ and $0.55$--$0.91$\,\Mpc\
across the four bins (\Cref{tab:sample}). Both confine the analysis to the
regime where the tree-level model has been shown to agree with
simulations~\cite{Gil-Marin:2014sta,Lazanu2015b,Hashimoto:2017klo,Chan2017,Oddo:2019run}.

We constrain the local, equilateral and orthogonal templates, defined in \cref{app:templates}. Neither the equilateral nor the
orthogonal template generates a scale-dependent bias, so for those two shapes
the power spectrum carries no $\fnl$ information and the constraint comes
entirely from the bispectrum.

Bispectrum forecasts usually approximate the triangle covariance by
its Gaussian diagonal part,
\begin{equation}
{\sf C}^{\rm G}_{ij}=\frac{(2\pi)^6}{V_{\rm s}V_{123}}\,s_{123}\,\delta_{ij}\,
P(\bk_1)P(\bk_2)P(\bk_3)\,,
\label{eq:covG}
\end{equation}
where $V_{123}=8\pi^2k_1k_2k_3(\Delta k)^3$ is the Fourier volume of the
triangle bin~\cite{Sefusatti2006} and $s_{123}=6,2,1$ for equilateral, isosceles
and scalene triangles. \Cref{eq:covG} treats every triangle as an independent
measurement. The non-Gaussian covariance instead couples triangle
configurations, most strongly in the squeezed limit. Neglecting it therefore
overcounts the information most severely for local PNG~\cite{
Barreira:2019icq,Biagetti:2021tua,Floss:2022wkq}, even at high redshift.
We therefore use the full bispectrum covariance,
${\sf C}_{ij}={\sf C}^{\rm G}_{ij}+{\sf C}^{\rm NG}_{ij}$, retaining its leading
non-Gaussian contribution,
\begin{multline}
{\sf C}^{\rm NG}_{ij}=\frac{2\,(2\pi)^3}{V_{\rm s}V_{123}^{i}V_{123}^{j}}
\Big[\delta_{k_1^ik_1^{\,j}}\,U(k_1^i,k_1^{\,j})\\
\times B(\bk_1^{\,j},\bk_2^{i},\bk_3^{i})\,
        B(\bk_1^{i},\bk_2^{\,j},\bk_3^{\,j})+8\,{\rm perm}\Big],
\label{eq:covNG}
\end{multline}
with the geometric factor $U$ given in \cref{app:cov}. \Cref{eq:covNG} is the
bispectrum--bispectrum term of Ref.~\cite{Barreira:2019icq}; the factor of
two approximates the trispectrum--power-spectrum term, which is of comparable
size for squeezed configurations~\cite{Biagetti:2021tua}. Its numerical
evaluation is described in \cref{app:binning}.

\section{Results}

\begin{figure*}[t]
\includegraphics[width=\textwidth]{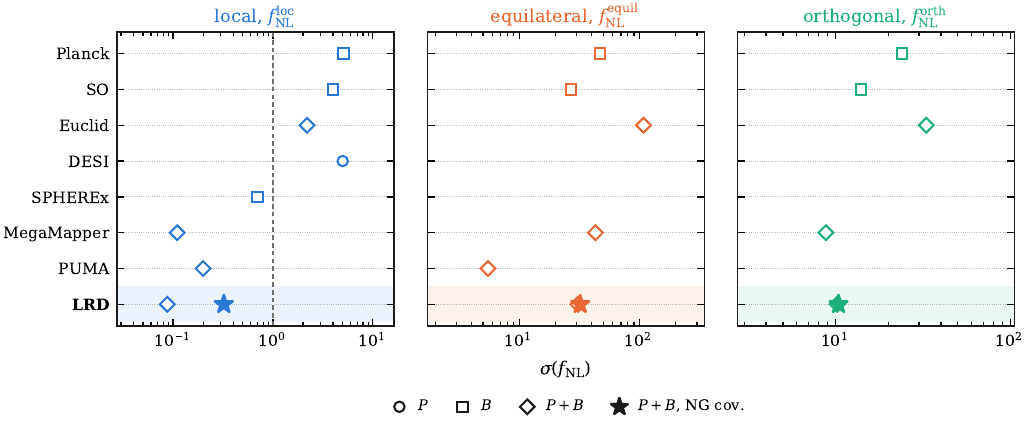}
\caption{Selected CMB and LSS forecasts, one panel per template. Literature
values are from Ref.~\cite{Achucarro_2022}, except the SPHEREx multi-tracer $B$
forecast~\cite{Heinrich:2023qaa}, the {\em Euclid}~\cite{Euclid:2025hlc} $P+B$ forecast and the Simons Observatory (SO)
baseline CMB bispectrum forecast, combined with {\em Planck}~\cite{SimonsObservatory:2018koc}. Circles denote $P$,
squares $B$ and diamonds $P+B$. The bottom row is our LRD forecast at the
conservative cut, with the Gaussian diagonal (open diamond) and the
non-Gaussian (filled star) covariance. The dashed line marks $\sigma(\fnl)=1$.
The templates are normalised differently, so each panel sets its own range.}
\label{fig:surveys}
\end{figure*}

\begin{table}[t]
\caption{Forecast $1\sigma$ errors on $\fnl$ for a $14{,}000\,{\rm deg}^2$ LRD
survey over $4<z<9$. Columns 3--5 use the Gaussian diagonal bispectrum
covariance; column 6 is the joint $P+B$ constraint with the non-Gaussian
covariance included (see text), whose conservative-cut entries are our main
results. The last column is the {\em Planck} 2018 bound~\cite{Planck:2019kim}. Neither
the equilateral nor the orthogonal template imprints a scale-dependent bias,
so $P$ alone carries no information for them. All errors are marginalised over
$\{\omega_b,\omega_c,h,n_s,\ln A_s\}$ and $32$ nuisance parameters.}
\label{tab:main}
\begin{ruledtabular}
\begin{tabular}{llccccc}
 & $\kmax$ & $P$ & $B$ & $P+B$ & $P+B$ & {\em Planck} \\
 &         &     &     & (diag.) & (NG cov.) &  \\
\hline
$\fnll$ & $0.50/\sigma_v$ & 0.385 & 0.090 & 0.088 & 0.322 & 5.10 \\
 & $0.75/\sigma_v$ & 0.383 & 0.053 & 0.052 & 0.301 &  \\
\hline
$\fnle$ & $0.50/\sigma_v$ & -- & 34.4 & 30.7 & 32.1 & 47.0 \\
 & $0.75/\sigma_v$ & -- & 28.9 & 26.5 & 28.4 &  \\
\hline
$\fnlo$ & $0.50/\sigma_v$ & -- & 10.5 & 10.2 & 10.4 & 24.0 \\
 & $0.75/\sigma_v$ & -- & 9.79 & 9.54 & 9.79 &  \\

\end{tabular}
\end{ruledtabular}
\end{table}

\Cref{tab:main} shows the forecast errors. Our main results, from the
joint $P+B$ analysis assuming non-Gaussian covariance and a conservative scale cut,
are
\begin{equation}
\sigma(\fnll)=0.32,\qquad \sigma(\fnle)=32,\qquad \sigma(\fnlo)=10,
\label{eq:headline}
\end{equation}
improving on {\em Planck} by factors of $16$, $1.5$ and $2.3$.

The effect of the non-Gaussian covariance, shown in the right panel of \Cref{fig:gridcov} and in
\Cref{tab:covariance}, depends strongly on the shape of the bispectrum. For equilateral and
orthogonal shapes the correction is $2$--$8\%$: for these templates the
non-Gaussian covariance remains close to diagonal. For the local shape, the
bispectrum-only constraint degrades by a factor of $5.9$
with a conservative scale cut and $9.0$ for an optimistic cut. The larger factor does not mean more information is lost: the Gaussian diagonal errors keep shrinking with $\kmax$ as added triangles are treated as independent, while the non-Gaussian errors barely change. The joint $P+B$
constraint degrades by less, factors of $3.7$ and $5.8$ respectively, because
the scale-dependent bias in the power spectrum is unaffected by the bispectrum covariance. With the Gaussian diagonal
covariance the bispectrum dominates the local constraint, while with the
non-Gaussian covariance the power spectrum and the bispectrum provide similar constraining power (\cref{app:cov}).

The covariance correction also makes the local result nearly insensitive to the scale cut. With the Gaussian diagonal
covariance, moving from $\kmax=0.50/\sigv$ to $0.75/\sigv$ tightens
$\sigma(\fnll)$ by a factor of $1.7$, while the non-Gaussian covariance treatment gains
only a factor of $1.07$. The extra squeezed triangles share their long modes
with those already included and so add little new information. In other words, the
squeezed information has saturated.

\Cref{fig:surveys} compares these results against published CMB and LSS
forecasts. The LSS forecasts shown adopt a Gaussian bispectrum covariance, so we
first compare them with our Gaussian diagonal results. On that footing, LRDs
give the tightest local bound of any forecast shown: $\sigma(\fnll)=0.088$ at
the conservative cut and $0.052$ at the optimistic cut, about $25$ times tighter
than {\em Euclid}, eight to thirteen times tighter than the SPHEREx bispectrum
forecast~\cite{Heinrich:2023qaa}\footnote{Including the power spectrum, SPHEREx
targets $\sigma(\fnll)\approx0.5$~\cite{Heinrich:2023qaa}.}, two to four times tighter than PUMA, and ahead
of the Stage-V MegaMapper concept, while covering less than half the
SPHEREx~\cite{Spherex_2015} sky area. For equilateral and orthogonal PNG, LRDs are about three times tighter
than {\em Euclid} and comparable to SO and MegaMapper, with only the PUMA
21\,cm concept forecasting a tighter equilateral bound. Including
non-Gaussian covariance changes the equilateral and orthogonal
constraints by only a few per cent. In contrast, it loosens the local bound to $\sigma(\fnll)=0.32$, which is still
about seven times tighter than {\em Euclid}, more than an order of magnitude
tighter than the SO CMB forecast, twice as tight as SPHEREx, and still comparable to PUMA. Differences
in survey modelling, nuisance parameters and covariance treatment make this an
indicative comparison. It nonetheless shows that a dedicated LRD survey could compete with, and even outperform, surveys designed specifically for PNG.

\section{Robustness}

\begin{figure*}[t]
\includegraphics[width=\textwidth]{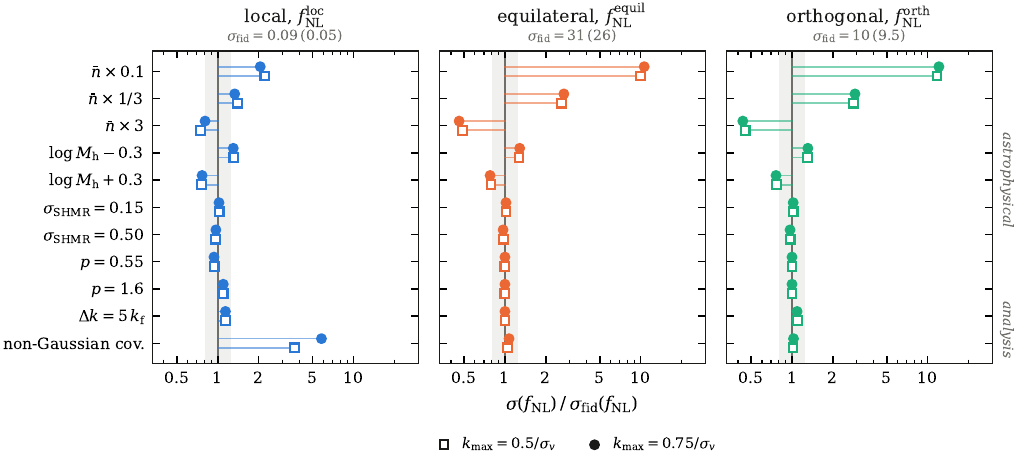}
\caption{Robustness tests for $P+B$ with the Gaussian diagonal covariance,
expressed as ratios to the fiducial forecast; the fiducial values
$\sigma_{\rm fid}$ are quoted in each panel for the conservative and (in
parentheses) optimistic cuts. Open squares use $\kmax=0.50/\sigv$, filled
circles the optimistic $\kmax=0.75/\sigv$; the shaded band marks $\pm20\%$. The
last two rows show the effect of coarser triangle binning and of the
non-Gaussian covariance (see \cref{app:binning}). The PNG bias response $p$ does not affect
the equilateral and orthogonal templates, which generate no scale-dependent
bias. The abundance dominates the astrophysical error budget; the equilateral
and orthogonal shapes, whose information sits on shot-noise-limited
small-scale triangles, are the most sensitive to it.}
\label{fig:robustness}
\end{figure*}

\Cref{fig:robustness} and \Cref{tab:robustness} summarise our robustness tests,
presenting the ratio between the constraints obtained using a modified survey/sample and the fiducial forecast. These ratios are obtained
with a Gaussian diagonal covariance.

{\em PNG bias response.} All LSS constraints on $\fnll$ are affected by a
degeneracy between $\fnll$ and the local-PNG bias
$b_\phi$~\cite{Afshordi2008,McDonald2008,Barreira:2021ueb,Barreira:2022sey}.
The latter depends on the halo formation history~\cite{Lazeyras:2022koc}, i.e.,\
on properties beyond total mass, so the universal relation
$b_\phi=2\delta_c(b_1-p)$ with $p=1$ need not hold for a selected tracer: LRDs
may be AGN- or merger-selected, in which case $p$ can depart substantially from
unity~\cite{Barreira:2020ekm,Barreira:2022sey}. If $b_\phi$ is accurately
modelled, e.g.\ with hydrodynamical simulations~\cite{Fondi:2023egm}, or
measured from the redshift evolution of the tracer number
counts~\cite{Nascimento2026}, the degeneracy can be broken. Here we instead span $p=0.55$ (a stellar-mass-selected value) to $p=1.6$ (recent-merger-like)~\cite{Slosar2008} and
find that $\sigma(\fnll)$ shifts by only $-6\%$ to $+10\%$. This is a
consequence of the high LRD bias: since $b_1\gg p$, varying $p$ changes
$b_\phi$ by at most $25\%$ in the lowest-redshift bin and by less than $10\%$
in the highest. The bispectrum, which dominates the local information, is even
less sensitive, since it constrains $\fnll$ also through $b_{\phi\delta}$,
which changes by less than $15\%$, and through the primordial bispectrum, which
does not depend on $p$ at all. With the non-Gaussian covariance, the power
spectrum carries a larger share of the local information, but its sensitivity
is still bounded by these modest changes in $b_\phi$.

{\em Abundance.} We next investigate the dominant astrophysical uncertainty:
the LRD abundance, whose model is uncertain at the level of a factor of a few
\cite{Inayoshi2025}. We rescale $\bar n$ by factors of $0.1$, $1/3$ and $3$.
The equilateral and orthogonal constraints, whose information sits on
shot-noise-limited small-scale triangles, are the most sensitive to the abundance,
degrading by factors of $10$--$12$ when $\bar n$ drops tenfold. The local
constraint is only affected by a factor of $2.0$--$2.2$, because its signal is
concentrated in squeezed triangles whose long side lies on large signal-dominated scales. For three times the fiducial abundance, the
local forecast improves by about $20\%$.

{\em Host mass and scatter.} Shifting the host halo mass by $\pm0.3$\,dex --
comfortably wider than the quoted uncertainty on the LRD host stellar
mass~\cite{sun2026little} and consistent with the spread in the direct clustering
estimates~\cite{lin2026large} -- degrades all three forecasts by $27$--$31\%$ for lower host masses and improves
them by $21$--$25\%$ for higher ones, as expected since more massive hosts are
more biased. Changing the lognormal
scatter of \Cref{eq:hod} from $0.30$\,dex to $0.15$ or $0.50$\,dex has a percent-level effect on the forecasts.

{\em Footprint.} All three constraints improve as the survey widens
(\Cref{fig:area_supp,tab:area}). Equilateral follows the mode-counting expectation,
$\sigma\propto A_{\rm sky}^{-0.50}$, while local improves faster,
$A_{\rm sky}^{-0.61}$, because the larger volume also lowers $\kf$ and opens more
squeezed configurations. Wide area is therefore especially valuable for local
PNG; orthogonal lies between these limits, scaling as $A_{\rm sky}^{-0.54}$.

\section{Discussion}

This analysis shows that a population first identified as compact red sources in JWST imaging could
become a leading probe of inflation. Our forecast, $\sigma(\fnll)=0.32$, lies a
factor of three below the $\sigma(\fnll)\simeq1$ threshold between single- and
multi-field inflation, while the equilateral and orthogonal constraints improve on
{\em Planck} by factors of $1.5$ and $2.3$. For local non-Gaussianity, similar constraints have only been forecast for hypothetical high-redshift 21\,cm surveys, where foregrounds were mostly neglected \cite{Karagiannis:2019jjx,PUMA_surv,Floss:2022wkq}. Furthermore, such constraints would require an interferometer with baselines of order $1$--$10$\,km operating from the far side of the Moon, shielded from radio-frequency interference and the Earth's ionosphere.\footnote{See, e.g., Figure~17 of \cite{Floss:2022grj}, which also neglects foregrounds and non-Gaussian covariance.} 

The properties of LRDs are close to ideal for a PNG analysis, and rarely
combined in a single tracer: they are abundant enough to keep shot noise under
control,\footnote{Over our $14{,}000\,{\rm deg}^2$ footprint the fiducial
abundance corresponds to $\approx2\times10^7$ LRDs, comparable to the
spectroscopic samples of DESI~\cite{DESI_DR2_II} and {\em Euclid}~\cite{Euclid:2024yrr}.} strongly biased, which amplifies both the clustering and the
scale-dependent-bias signal, and they fill a vast volume at $4<z<9$, where short
modes remain perturbative. Their large volume is particularly
important for local PNG because it exposes the long modes that anchor squeezed
triangles.

The forecast is also stable against several uncertain ingredients. Broad changes
in the PNG response, the host mass, occupation scatter and small-scale cut
leave the main results intact. The LRD abundance is the main astrophysical
uncertainty: a tenfold reduction weakens the local constraint by only about a
factor of two, although it severely reduces the equilateral and orthogonal
constraints.

The $\sigma(\fnle)\sim1$ regime~\cite{Meerburg:2019qqi,Achucarro_2022} remains challenging: combining the abundance and area scalings of \cref{fig:robustness,fig:area_supp}, even a full-sky survey with three times our fiducial $\bar n$ would yield $\sigma(\fnle)\approx7$--$9$, nearly an order of magnitude tighter than {\em Planck}, yet still short of the target. Reaching it with LRDs would require extracting information from smaller, nonlinear scales, through improved modelling or simulation-based analyses~\cite{Jung:2022gfa,Coulton:2022rir}, by exploiting the non-locality of the primordial signal~\cite{Baumann:2021ykm}, or by using statistics such as $k$-nearest-neighbour distributions~\cite{Coulton:2023ouk} and the density PDF~\cite{Friedrich:2019byw}. For comparison, the lunar far-side interferometer discussed above would need baselines of order $10$\,km to reach this equilateral precision~\cite{Floss:2022grj}.

The non-Gaussian bispectrum covariance is essential for local PNG, and our
treatment follows approximations shown to be adequate for this shape~\cite{Barreira:2019icq,Biagetti:2021tua}. Since it saturates the local information, further gains in $\sigma(\fnll)$ must come mainly from larger volumes and from priors on $b_\phi$, rather than from smaller scales. A realistic analysis must also model relativistic
projection and wide-angle effects~\cite{Umeh:2016nuh,Clarkson:2018dwn,Maartens:2019yhx,Maartens:2020jzf,Noorikuhani:2022bwc,Pardede:2023ddq}, which are largest on the scales where the
local signal peaks, and control spurious large-scale power
from the target selection.

The survey we presented here does not exist and has not been proposed~\cite{Zebrowski2026}. LRD redshifts come most efficiently from
H$\alpha$, which across $4<z<9$ is redshifted to $3.3$--$6.6\,\mu$m. That
single requirement excludes every wide-field spectroscopic facility now
operating or approved: {\em Euclid} and Roman slitless spectroscopy both stop below
$2.2\,\mu$m, and SPHEREx, although it reaches $5\,\mu$m, lacks the sensitivity
for these sources. JWST has the wavelength coverage and the sensitivity but not
the survey speed -- its largest contiguous extragalactic fields span under a
square degree~\cite{Casey23}, four orders of magnitude short of the $10^4\,{\rm
deg}^2$ assumed here. What is needed is therefore not a deeper telescope but a
faster one: a wide-field, highly multiplexed near- to mid-infrared
spectrograph targeting LRDs pre-selected by their distinctive
colours~\cite{Hviding25,Setton25,Matthee26}. LRD PNG is therefore a science case
for such a facility, in the spirit of the Stage-V concepts now being
developed~\cite{schlegel2019astro2020,schlegel2022megamapper,
besuner2025spectroscopic} but pushed redward, with the potential to place
stringent constraints on inflation.

\begin{acknowledgments}
DK and PDM thank the Friday Archive Coffee discussion for spearheading this idea. DK is supported by the Dutch Research Council (NWO) (grant no. OCENW.M.22.307). DK and PDM like to thank the Center for Information Technology of the University of Groningen for their support and for providing access to the Hábrók high-performance computing cluster where the entire analysis was run. We acknowledge the use of Claude agents for help with writing and editing the draft, plotting, and workflow management with the analysis software developed by DK. The authors directed and checked all outputs used here and retain full responsibility for the scientific content. The analysis software used in this paper is stored in a private Git repository and is available upon reasonable request. All assumptions in the code are provided in the appendix. 
\end{acknowledgments}

\bibliographystyle{apsrev4-2}
\bibliography{references}

\appendix
\crefalias{section}{appendix}   
\onecolumngrid

\section{Primordial bispectrum templates}
\label{app:templates}

We consider the three most studied shapes of the primordial bispectrum, each
peaking at a different triangle configuration and pointing to a different
departure from single-field slow-roll inflation: {\em
local}~\cite{Salopek1990,Gangui1993,Verde1999,Komatsu2001}, peaking on squeezed
triangles ($k_3\ll k_2\simeq k_1$) and generated by additional light fields;
{\em equilateral}~\cite{Creminelli2005}, peaking on equilateral configurations
($k_1\simeq k_2\simeq k_3$) and generated by non-trivial inflaton
self-interactions; and {\em orthogonal}~\cite{Chen:2006nt,Senatore2009}, peaking on both
equilateral and folded ($k_1\simeq k_2\simeq k_3/2$) triangles. For the
orthogonal shape we use the {\em orthogonal-LSS} approximation of
Ref.~\cite{Senatore2009}, which is the one relevant for large-scale structure
because it has a regular squeezed limit. In terms of the Bardeen
gauge-invariant primordial potential $\Phi$, the templates
are~\cite{Euclid:2025hlc}
\begin{align}
B_{\Phi}^{\rm loc}&=2\fnll\Big[P_{\Phi}(k_1)P_{\Phi}(k_2)+2\ {\rm perm}\Big]\,,
\label{eq:Bloc}\\[2pt]
B_{\Phi}^{\rm equil}&=6\fnle\Big(
 -\big[P_{\Phi}(k_1)P_{\Phi}(k_2)+2\ {\rm perm}\big]
 -2\big[P_{\Phi}(k_1)P_{\Phi}(k_2)P_{\Phi}(k_3)\big]^{2/3} \nonumber\\
&\qquad\quad +\big[P_{\Phi}^{1/3}(k_1)P_{\Phi}^{2/3}(k_2)P_{\Phi}(k_3)
 +5\ {\rm perm}\big]\Big)\,,
\label{eq:Bequil}\\[2pt]
B_{\Phi}^{\rm orth}&=6\fnlo
 \big[P_\Phi(k_1)P_\Phi(k_2)P_\Phi(k_3)\big]^{2/3}
 \left[(1+q)\frac{\Delta}{k_1k_2k_3}-q\frac{\Gamma^3}{k_1^2k_2^2k_3^2}\right],
\label{eq:Borth}
\end{align}
with
\begin{equation}
q=\frac{27}{-21+\dfrac{743}{7(20\pi^2-193)}}\,,\quad
\Delta=\prod_{i=1}^{3}(k_T-2k_i)\,,\quad
\Gamma=\tfrac23(k_1k_2+k_2k_3+k_3k_1)-\tfrac13(k_1^2+k_2^2+k_3^2)\,,
\end{equation}
and $k_T=k_1+k_2+k_3$. The potential is related to the linear matter field by
$\delta^{\rm L}_m(\bk,z)=\mathcal{M}(k,z)\Phi(\bk)$, so that
$P_\Phi(k)=P^{\rm L}_m(k,z)/\mathcal{M}^2(k,z)$, where
$\mathcal{M}(k,z)=2c^2D(z)T(k)k^2/(3\Omega_mH_0^2 g_{\rm dec})$ contains the
growth factor $D(z)$ and the matter transfer function $T(k)$, with $g_{\rm dec}$
the Bardeen-potential growth factor at decoupling, ensuring that $\fnl$ is in the
CMB convention~\cite{Camera:2014bwa}. The primordial contribution to the matter bispectrum is
$B_I=\mathcal{M}(k_1)\mathcal{M}(k_2)\mathcal{M}(k_3)B_\Phi$. 

The squeezed behaviour of the template also fixes which statistics carry
information. Only the local shape produces a scale-dependent bias; the
equilateral and orthogonal templates approach the squeezed limit fast enough
that the correction becomes scale-independent on large scales and is absorbed
into $b_1$~\cite{Schmidt2010,Scoccimarro2011,Assassi2015}. For those two shapes,
the power spectrum therefore carries no $\fnl$ information, and the constraint
comes entirely from the primordial term in the bispectrum.

\section{Tree-level model}
\label{app:model}

The observed redshift-space power spectrum and bispectrum of the LRD sample are
modelled at tree level as
\begin{align}
P(\bk,z)&=\mathcal{A}_P\,D^P(\bk)\,Z_1^2(\bk)\,P^{\rm L}_m(k,z)
          +P_\varepsilon(z)\,,
\label{eq:Pmodel}\\[2pt]
B(\bk_1,\bk_2,\bk_3,z)&=\mathcal{A}_B\Big\{
 D^B(\bk_1,\bk_2,\bk_3)\Big[Z_1(\bk_1)Z_1(\bk_2)Z_1(\bk_3)\,
 B_I(k_1,k_2,k_3) \nonumber\\
&\qquad +2\,Z_1(\bk_1)Z_1(\bk_2)Z_2(\bk_1,\bk_2)P^{\rm L}_m(k_1)P^{\rm L}_m(k_2)
 +2\ {\rm perm}\Big] \nonumber\\
&\qquad +2P_{\varepsilon\varepsilon_\delta}(z)
 \Big[Z_1(\bk_1)P^{\rm L}_m(k_1)+2\ {\rm perm}\Big]
 +B_\varepsilon(z)\Big\}\,,
\label{eq:Bmodel}
\end{align}
where $P^{\rm L}_m$ is the linear matter power spectrum from
CAMB~\cite{CAMB}, infrared-resummed so that the acoustic feature is damped
consistently~\cite{Assassi2015,Ivanov:2021kcd}. The Alcock--Paczy\'nski
factors~\cite{Alcock1979,Seo2003,Song2015} are
$\mathcal{A}_P=(D_A^{\rm fid}/D_A)^2(H/H^{\rm fid})$ and
$\mathcal{A}_B=(D_A^{\rm fid}/D_A)^4(H/H^{\rm fid})^2$, and the wavenumbers and
angles entering \Cref{eq:Pmodel,eq:Bmodel} are the ones mapped by the same
distortion. The damping factors combine the Fingers-of-God
effect~\cite{Jackson1972,Peacock1994,Ballinger1996} with the spectroscopic
redshift error,
\begin{align}
D^P(\bk)&=\exp\!\Big[-k^2\mu^2\big(\tfrac{1}{2}\sigma_P^2+\sigma_r^2\big)\Big],\\
D^B(\bk_1,\bk_2,\bk_3)&=\exp\!\Big[-\tfrac{1}{2}\big(\sigma_B^2+\sigma_r^2\big)
\sum_{i=1}^{3}k_i^2\mu_i^2\Big],
\end{align}
with $\sigma_r=c\,(1+z)\,[\sigma_z/(1+z)]/H(z)$ the radial smearing from the
redshift error, and the damping scales fixed at fiducial to
$\sigma_P=\sigma_B=\sqrt2\,f(z)\,\sigv(z)$, where
$\sigv^2(z)=\int{\rm d}k\,P^{\rm L}_m(k,z)/(6\pi^2)$. Both are varied and
marginalised over. The stochastic amplitudes take the Poisson
values~\cite{Schmidt2015,Desjacques2016} $P_\varepsilon=1/\bar n$,
$P_{\varepsilon\varepsilon_\delta}=b_1/(2\bar n)$ and
$B_\varepsilon=1/\bar n^2$ at fiducial, and are likewise free parameters.

The redshift-space kernels, including local PNG, are~\cite{Baldauf2011,Tellarini2016,Karagiannis2018}
\begin{align}
Z_1(\bk_i)&=b_1+f\mu_i^2+\frac{\fnl\,b_\phi}{\mathcal{M}(k_i,z)}\,,
\label{eq:Z1}\\
Z_2(\bk_i,\bk_j)&=b_1F_2(\bk_i,\bk_j)+f\mu_{ij}^2G_2(\bk_i,\bk_j)
 +\frac{b_2}{2}+\frac{b_{s^2}}{2}S_2(\bk_i,\bk_j)
 +\frac{f\mu_{ij}k_{ij}}{2}
 \left[\frac{\mu_i}{k_i}Z_1(\bk_j)+\frac{\mu_j}{k_j}Z_1(\bk_i)\right]
 \nonumber\\
&\quad +\fnl\left\{\frac{b_{\phi\delta}}{2}\left[\frac{1}{\mathcal{M}(k_i,z)}
 +\frac{1}{\mathcal{M}(k_j,z)}\right]-b_\phi\,N_2(\bk_i,\bk_j)\right\},
\label{eq:Z2}
\end{align}
where $f$ is the growth rate, $\mu_i=\hat\bk_i\cdot\hat{\bm z}$,
$\mu_{ij}=(\mu_ik_i+\mu_jk_j)/k_{ij}$, $k_{ij}^2=(\bk_i+\bk_j)^2$ and
$\mathcal{M}$ is defined in \cref{app:templates}. $F_2$ and $G_2$ are the second-order
symmetric SPT kernels~\cite{Bernardeau2002},
$S_2(\bk_1,\bk_2)=(\hat\bk_1\cdot\hat\bk_2)^2-1/3$ is the tidal
kernel~\cite{McDonald2009,Baldauf2012}, and
$N_2(\bk_i,\bk_j)=\tfrac{1}{2}\,(\bk_i\cdot\bk_j)\big[k_j^{-2}\mathcal{M}^{-1}(k_i,z)
+k_i^{-2}\mathcal{M}^{-1}(k_j,z)\big]$
is the symmetrised kernel that encodes the coupling of the PNG
potential to the Eulerian-to-Lagrangian displacement
field~\cite{Giannantonio2010,Baldauf2011}. For the equilateral and
orthogonal templates, the squeezed limit is regular, the scale-dependent bias
term in \Cref{eq:Z1} vanishes, and the $\fnl$ information comes entirely from the
primordial term in \Cref{eq:Bmodel}.

The full parameter vector is
\begin{align}
\bm\theta=\Big\{&\omega_b,\omega_c,h,n_s,\ln A_s,\fnl;\ \nonumber\\
&D_A(z_i),H(z_i),f(z_i),b_1(z_i),b_2(z_i),b_{s^2}(z_i),
\sigma_P(z_i),\sigma_B(z_i),
P_\varepsilon(z_i),P_{\varepsilon\varepsilon_\delta}(z_i),B_\varepsilon(z_i)
\Big\}\,,
\label{eq:params}
\end{align}
with $i=1\ldots4$. The per-bin geometric and growth parameters are projected onto
the cosmological basis before marginalisation~\cite{Euclid:2019clj}, leaving the
eight bias, velocity-dispersion and stochastic parameters per bin -- $32$ in
total -- to be marginalised directly. Fiducial cosmological values are those of
{\em Planck} 2018~\cite{Planck2018}.

\section{Bias model}
\label{app:bias}

\begin{figure}[h]
\centering
\includegraphics[width=0.42\textwidth]{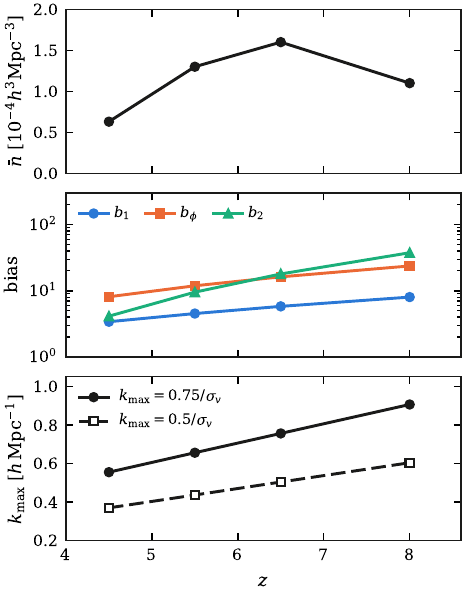}
\caption{Per-bin inputs to the LRD forecast. {\em Top}: comoving number density
from the abundance model of Ref.~\cite{Inayoshi2025}. {\em Middle}: the bias
hierarchy generated by the SHMR-anchored occupation of \Cref{eq:hod} --- linear
bias $b_1$, local-PNG response $b_\phi$, and quadratic bias $b_2$. {\em Bottom}:
the two small-scale cuts, $\kmax(z)=\alpha/\sigv(z)$ with $\alpha=0.50$ and
$0.75$.}
\label{fig:survey}
\end{figure}

LRDs are modelled as central objects with a mean occupation that is lognormal in
host mass,
\begin{equation}
\langle N_{\rm c}(\Mh)\rangle
= \mathcal{A}\,
  \exp\!\left[-\frac{\big(\log_{10}\Mh-\log_{10}M_{\rm c}\big)^2}
                    {2\,\sigma_{\rm SHMR}^2}\right],
\label{eq:hod}
\end{equation}
of width $\sigma_{\rm SHMR}=0.30$\,dex. The centre $M_{\rm c}$ is fixed by
requiring that the mean host mass of {\em occupied} haloes reproduce the SHMR
host mass of the main text, and the duty cycle $\mathcal{A}$ is then set by the
tabulated abundance, so that the SHMR controls the mass -- and hence bias --
distribution while the abundance controls only the occupied fraction.
\Cref{eq:hod} is convolved with the Tinker mass function and
halo bias~\cite{Tinker2008,tinker2010large} to give the effective tracer biases,
\begin{equation}
b_X=\frac{1}{\bar n}\int{\rm d}\Mh\,\frac{{\rm d}n}{{\rm d}\Mh}\,
\langle N_{\rm c}(\Mh)\rangle\,b_X^{\rm h}(\Mh,z)\,,
\qquad
\bar n=\int{\rm d}\Mh\,\frac{{\rm d}n}{{\rm d}\Mh}\,\langle N_{\rm c}(\Mh)\rangle\,,
\label{eq:beff}
\end{equation}
for $X=1,2$, with the tidal bias fixed by co-evolution,
$b_{s^2}=-\tfrac{4}{7}(b_1-1)$~\cite{Baldauf2012,Chan2012}. The PNG responses are
obtained from the same integral with the halo-level
responses~\cite{Slosar2008,Desjacques2016,Desjacques2009}
\begin{align}
b_\phi^{\rm h}&=2\delta_c\,(b_1^{\rm h}-p)\,,\\
b_{\phi\delta}^{\rm h}&=b_\phi^{\rm h}
 +2\delta_c\left[b_2^{\rm h}-\tfrac{8}{21}\big(b_1^{\rm h}-1\big)\right]
 -6\,\big(b_1^{\rm h}-1\big)\,,
\end{align}
where $\delta_c=1.686$ and $p$ parametrises the departure from a universal mass
function; $p=1$ is our fiducial choice, and the range $p=0.55$--$1.6$ is explored
in the main text. Because the occupation centre is solved for the mean host mass
of occupied haloes -- not for a target $b_1$ -- the linear bias is a prediction of
the model rather than an input, which is what makes the $1\%$ agreement with
Ref.~\cite{Zebrowski2026} a meaningful check. The per-bin inputs are collected in
\Cref{fig:survey}.

\begin{table}[h]
\caption{Per-bin properties of the LRD sample. The number densities are from the
abundance model of Ref.~\cite{Inayoshi2025}; the host mass is the SHMR prediction
at the bin centre; the biases are the effective tracer biases predicted
by the occupation model of \Cref{eq:hod}, with $p=1$ for the PNG responses; and
the two small-scale cuts are $\kmax(z)=\alpha/\sigv(z)$.}
\label{tab:sample}
\begin{ruledtabular}
\begin{tabular}{lcccc}
 & bin 1 & bin 2 & bin 3 & bin 4 \\
\hline
$z$ range & $4$--$5$ & $5$--$6$ & $6$--$7$ & $7$--$9$ \\
$\bar n\,[10^{-4}\,h^3{\rm Mpc}^{-3}]$ & 0.63 & 1.3 & 1.6 & 1.1 \\
$V_{\rm s}\,[(\mathrm{Gpc}/h)^3]$ & 46.7 & 41.7 & 37.3 & 63.8 \\
$\kf\,[10^{-3}\,h\,\mathrm{Mpc}^{-1}]$ & 1.74 & 1.81 & 1.88 & 1.57 \\
$\log_{10}(\Mh/M_\odot)$ & 11.00 & 10.94 & 10.88 & 10.78 \\
$b_1$ & 3.41 & 4.53 & 5.81 & 8.03 \\
$b_2$ & 4.13 & 9.53 & 17.9 & 37.5 \\
$b_{s^2}$ & $-1.38$ & $-2.02$ & $-2.75$ & $-4.02$ \\
$b_\phi$ & 8.12 & 11.9 & 16.2 & 23.7 \\
$b_{\phi\delta}$ & 14.2 & 32.5 & 61.0 & 127.2 \\
$\sigv\,[\mathrm{Mpc}/h]$ & 1.35 & 1.14 & 0.99 & 0.83 \\
$\kmax(\alpha=0.50)\,[h\,\mathrm{Mpc}^{-1}]$ & 0.37 & 0.44 & 0.50 & 0.60 \\
$\kmax(\alpha=0.75)\,[h\,\mathrm{Mpc}^{-1}]$ & 0.55 & 0.66 & 0.76 & 0.91 \\

\end{tabular}
\end{ruledtabular}
\end{table}

\section{Covariance}
\label{app:cov}

The geometric factor appearing in \Cref{eq:covNG} is
$U(k_1^i,k_1^{\,j})=16\pi^3k_2^ik_3^ik_2^{\,j}k_3^{\,j}(\Delta k)^5$, and the
nine permutations run over the possible mode identifications between triangles
$i$ and $j$~\cite{Barreira:2019icq,Biagetti:2021tua,Salvalaggio:2024vmx}.
For flattened ($k_1=k_2+k_3$) and open configurations the thin-shell expressions
for $V_{123}$ and $U$ break down, and we use the analytic corrections of
Ref.~\cite{Biagetti:2021tua}. The matrix
${\sf C}={\sf C}^{\rm G}+{\sf C}^{\rm NG}$ is inverted by Cholesky
decomposition. The shot-noise contributions enter the Gaussian part of \Cref{eq:covG}
through the observed $P(\bk)$, and the bispectra in \Cref{eq:covNG} are the
tree-level model of \Cref{eq:Bmodel}.

Neglecting the power spectrum--bispectrum cross-covariance makes $P+B$ a sum of
two information matrices. With the Gaussian diagonal covariance, $B$ dominates
the local constraint ($0.053$ against $0.383$ for $P$ at the optimistic cut);
with the non-Gaussian covariance, the two become comparable and $P+B$ is $1.6$
times tighter than $B$ alone. This is the regime in which the cross-covariance
could matter most, and the one addressed by the local-PNG studies of
Refs.~\cite{Barreira:2019icq,Biagetti:2021tua}, which find its impact to be
minor over the scales considered here.

\section{Triangle binning}
\label{app:binning}

\begin{figure}[h]
\centering
\includegraphics[width=0.62\textwidth]{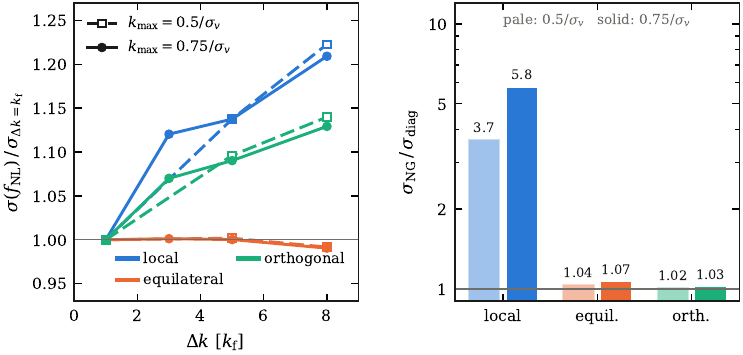}
\caption{{\em Left}: convergence with triangle-bin width, normalised to the
production grid $\Delta k=\kf$. The equilateral template is converged at any
binning; local and orthogonal are under-resolved by coarse bins because the
squeezed legs are smeared out. {\em Right}: ratio of the non-Gaussian to the
Gaussian diagonal forecast for $P+B$, measured on matched $\Delta k=8\,\kf$
grids. Only the local shape, which concentrates its information in squeezed
configurations, is strongly affected.}
\label{fig:gridcov}
\end{figure}

All forecasts in the main text bin the power spectrum and the triangles with a
width equal to the fundamental mode of each redshift bin, $\Delta k=\kf$
(\Cref{tab:sample}), with the upper limit of the triangle sum rounded down to
the nearest multiple of $\Delta k$; this gives the $\kmax$ values of
\Cref{tab:sample}. Such fine binning is needed to resolve the squeezed
configurations. \Cref{tab:convergence} and the left panel of \Cref{fig:gridcov}
show that coarser grids weaken the local and orthogonal constraints by $14\%$
and $9\%$ at $\Delta k=5\,\kf$, while the equilateral constraint is converged at
any binning.

Two sets of results are too expensive to compute at $\Delta k=\kf$ and are
instead expressed as ratios between two runs on an identical, coarser grid, with
the same $\kmax$ and parameter set, so that the binning cancels. The robustness
tests of \Cref{fig:robustness} and \Cref{tab:robustness} are computed on a
$\Delta k=5\,\kf$ grid. The non-Gaussian covariance is dense, so its cost scales
as $n_{\rm tri}^2$ and the exact treatment is affordable only at
$\Delta k=8\,\kf$, which gives $5.9\times10^4$ triangles at the optimistic cut;
a direct $\Delta k=\kf$ evaluation, with $n_{\rm tri}\sim10^7$, is out of reach.
Each non-Gaussian run is therefore paired with a Gaussian diagonal run on the
same grid (\Cref{tab:covariance}), and the ratio of the two is applied to the
Gaussian diagonal $\Delta k=\kf$ forecast to obtain \Cref{eq:headline}.

This transfer assumes that the covariance correction does not depend on the bin
width. We tested this at the conservative cut by repeating the paired runs on a
$\Delta k=5\,\kf$ grid (the optimistic cut at $5\,\kf$ would require a
$\sim140$\,GB dense matrix). For local, the ratio is $5.91$ for $B$ and $3.54$
for $P+B$, against $5.94$ and $3.68$ at $8\,\kf$; for equilateral and orthogonal
the $P+B$ ratios, $1.045$ and $1.016$, are unchanged. The correction therefore
does not grow as the grid is refined: the local ratios decrease slightly when it is refined
from $8\,\kf$ to $5\,\kf$, from $5.94$ to $5.91$ for $B$ and from $3.68$ to $3.54$ for $P+B$.

\begin{table}[h]
\caption{Convergence of $\sigma(\fnl)$ from $P+B$ with the triangle-bin width
$\Delta k$, in units of the fundamental mode $\kf$, with the Gaussian diagonal
covariance and the SHMR bias model. Blank entries were not run.}
\label{tab:convergence}
\begin{ruledtabular}
\begin{tabular}{llcccc}
 & $\kmax$ & $\Delta k=\kf$ & $3\,\kf$ & $5\,\kf$ & $8\,\kf$ \\
\hline
$\fnll$ & $0.50/\sigma_v$ & 0.088 & -- & 0.100 & 0.107 \\
 & $0.75/\sigma_v$ & 0.052 & 0.058 & 0.059 & 0.063 \\
$\fnle$ & $0.50/\sigma_v$ & 30.7 & -- & 30.8 & 30.5 \\
 & $0.75/\sigma_v$ & 26.5 & 26.5 & 26.5 & 26.2 \\
$\fnlo$ & $0.50/\sigma_v$ & 10.2 & -- & 11.2 & 11.6 \\
 & $0.75/\sigma_v$ & 9.54 & 10.2 & 10.4 & 10.8 \\

\end{tabular}
\end{ruledtabular}
\end{table}

\begin{table}[h]
\caption{The non-Gaussian covariance test. Each ``NG'' run is paired with a
Gaussian diagonal run on an identical $\Delta k=8\,\kf$ grid, so that the ratio isolates
the covariance treatment from every other choice.}
\label{tab:covariance}
\begin{ruledtabular}
\begin{tabular}{llcccccc}
 & $\kmax$ & $B$ diag. & $B$ NG & ratio & $P{+}B$ diag. & $P{+}B$ NG & ratio \\
\hline
$\fnll$ & $0.50/\sigma_v$ & 0.110 & 0.655 & 5.94 & 0.107 & 0.394 & 3.68 \\
 & $0.75/\sigma_v$ & 0.064 & 0.573 & 9.00 & 0.063 & 0.364 & 5.80 \\
$\fnle$ & $0.50/\sigma_v$ & 34.1 & 36.8 & 1.08 & 30.5 & 31.8 & 1.04 \\
 & $0.75/\sigma_v$ & 28.7 & 31.0 & 1.08 & 26.2 & 28.2 & 1.07 \\
$\fnlo$ & $0.50/\sigma_v$ & 12.2 & 12.3 & 1.01 & 11.6 & 11.8 & 1.02 \\
 & $0.75/\sigma_v$ & 11.2 & 11.5 & 1.03 & 10.8 & 11.1 & 1.03 \\

\end{tabular}
\end{ruledtabular}
\end{table}

\begin{table}[h]
\caption{Astrophysical robustness: $\sigma(\fnl)$ for each variation divided by
the fiducial value, computed with the Gaussian diagonal covariance on a matched $\Delta k=5\,\kf$ grid so that the
binning cancels. Each pair of columns gives the conservative ($0.50/\sigv$) and
optimistic ($0.75/\sigv$) scale cuts.}
\label{tab:robustness}
\begin{ruledtabular}
\begin{tabular}{llcccccc}
 & & \multicolumn{2}{c}{local} & \multicolumn{2}{c}{equilateral} & \multicolumn{2}{c}{orthogonal} \\
\hline
PNG bias response & $p=0.55$ & 0.94 & 0.94 & 1.00 & 1.00 & 1.00 & 1.00 \\
PNG bias response & $p=1.6$ & 1.09 & 1.10 & 1.00 & 1.00 & 1.00 & 1.00 \\
Abundance & $\bar n\times0.1$ & 2.22 & 2.05 & 10.02 & 10.62 & 11.70 & 12.09 \\
Abundance & $\bar n\times1/3$ & 1.40 & 1.33 & 2.62 & 2.72 & 2.85 & 2.91 \\
Abundance & $\bar n\times3$ & 0.75 & 0.81 & 0.49 & 0.46 & 0.45 & 0.43 \\
Host mass & $\log M_h-0.3$ & 1.31 & 1.30 & 1.27 & 1.28 & 1.30 & 1.31 \\
Host mass & $\log M_h+0.3$ & 0.75 & 0.77 & 0.79 & 0.78 & 0.77 & 0.76 \\
Host mass scatter & $\sigma_{\rm SHMR}=0.15$ & 1.02 & 1.02 & 1.02 & 1.02 & 1.02 & 1.02 \\
Host mass scatter & $\sigma_{\rm SHMR}=0.50$ & 0.96 & 0.97 & 0.97 & 0.97 & 0.97 & 0.97 \\

\end{tabular}
\end{ruledtabular}
\end{table}

\begin{table}[h]
\caption{Footprint scan for $P+B$ with the Gaussian diagonal covariance, and the
fitted power-law slope $\sigma\propto A_{\rm sky}^{-\beta}$. Areas are in
${\rm deg}^2$.}
\label{tab:area}
\begin{ruledtabular}
\begin{tabular}{llccccc}
 & $\kmax$ & $10{,}000$ & $14{,}000$ & $20{,}000$ & $30{,}000$ & $\beta$ \\
\hline
$\fnll$ & $0.50/\sigma_v$ & 0.124 & 0.099 & 0.080 & 0.062 & 0.62 \\
 & $0.75/\sigma_v$ & 0.072 & 0.058 & 0.047 & 0.037 & 0.61 \\
$\fnle$ & $0.50/\sigma_v$ & 36.1 & 30.7 & 25.6 & 20.8 & 0.50 \\
 & $0.75/\sigma_v$ & 31.0 & 26.4 & 22.1 & 18.0 & 0.50 \\
$\fnlo$ & $0.50/\sigma_v$ & 13.4 & 11.2 & 9.18 & 7.36 & 0.55 \\
 & $0.75/\sigma_v$ & 12.5 & 10.4 & 8.53 & 6.85 & 0.54 \\

\end{tabular}
\end{ruledtabular}
\end{table}

\begin{figure*}[h]
\includegraphics[width=\textwidth]{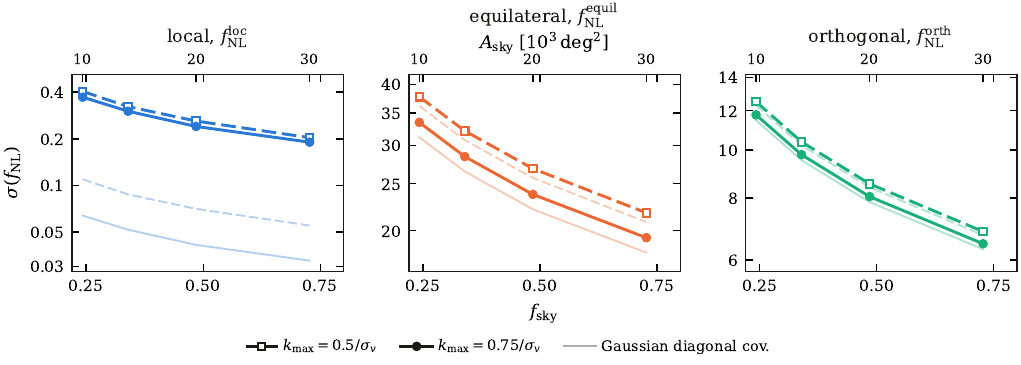}
\caption{Forecast $\sigma(\fnl)$ against sky fraction, with the corresponding
area on the upper axis. Heavy curves include the non-Gaussian bispectrum
covariance; faint curves use the Gaussian diagonal approximation. Dashed lines
with open squares use $\kmax=0.50/\sigv$; solid lines with filled circles use
$\kmax=0.75/\sigv$. The fitted
scalings are $A_{\rm sky}^{-0.61}$, $A_{\rm sky}^{-0.50}$ and
$A_{\rm sky}^{-0.54}$ for local, equilateral and orthogonal.}
\label{fig:area_supp}
\end{figure*}

\end{document}